\documentstyle[12pt,aaspp4]{article}

\def\bibi{\bibitem}

\def\beq{\begin{equation}}
\def\enq{\end{equation}}
\def\begar{\begin{eqnarray}}
\def\endar{\end{eqnarray}}
\def\Mesz{M\'esz\'aros}
\def\Pacz{Paczy\'nski}
\def\mathnew{\mathsurround=0pt}
\def\simov#1#2{\lower .5pt\vbox{\baselineskip0pt \lineskip-.5pt
        \ialign{$\mathnew#1\hfil##\hfil$\crcr#2\crcr\sim\crcr}}}
\def\simg{\mathrel{\mathpalette\simov >}}
\def\siml{\mathrel{\mathpalette\simov <}}
\def\etal{{\it et.al.}}

\received{31 November, 1995 }
\accepted{31 January, 1996}
\slugcomment{to appear in Astrophys. J. (Letters)}
 
\begin{document}
\title{
ELECTRON ACCELERATION AND EFFICIENCY IN \\
NONTHERMAL GAMMA RAY SOURCES
}


\author{A.M. Bykov\altaffilmark{1}}
\affil{A.F. Ioffe Physical-Technical Institute,  St. Petersburg, Russia, 194021}

\and
 
\author{P. \Mesz\altaffilmark{1,2}}
\affil{525 Davey Laboratory, Pennsylvania State University, University Park,
 PA 16803}
 
\altaffiltext{1}{Institute for Theoretical Physics, University of California,
  Santa Barbara, CA 93106}
\altaffiltext{2}{Center for Gravitational Physics and Geometry, Pennsylvania
  State University}

\begin{abstract}

In energetic nonthermal sources such as gamma-ray bursts, AGN
or galactic jet sources, etc., one expects both relativistic and 
transrelativistic shocks acompanied by violent motions of moderately
relativistic plasma. We present general considerations indicating
that these sites are electron and positron accelerators leading to a modified 
power law spectrum. The electron (or $e^\pm$) energy index is very hard, 
$\propto \gamma^{-1}$ or flatter up to a comoving frame break energy 
$\gamma_\ast$, and becomes steeper above that. In the example of gamma-ray 
bursts the Lorentz factor reaches $\gamma_\ast\sim 10^3$ for $e^{\pm}$ 
accelerated by the internal shock ensemble on subhydrodynamical time scales.  
For pairs accelerated on hydrodynamical timescales in the external shocks 
similarly hard spectra are obtained, and the break Lorentz factor can be as 
high as $\gamma_\star \siml 10^5$. 
Radiation from the nonthermal electrons produces photon spectra with shape 
and characteristic energies in qualitative agreement with observed generic 
gamma-ray burst and blazar spectra. The scenario described here provides a 
plausible way to solve one of the crucial problems of nonthermal high energy
sources, namely the efficient transfer of energy from the proton flow to an 
apropriate nonthermal lepton component.
\end{abstract}
 
\keywords{acceleration of particles-- shock waves -- cosmic rays: general -- 
gamma-rays: bursts -- gamma-rays: theory-- galaxies: active--galaxies:jets}
 
\section{Introduction}

Shocks, and more generally systems of shocks as well as turbulent flow 
downstream from such shocks may be a common feature of a number of nonthermal 
gamma-ray sources. A recent example of interest is the dissipative 
relativistic fireball model of gamma-ray bursts (GRB), which follows
from very general energetic and observational constraints,
independently of whether the sorces are at cosmological or galactic halo 
distances. This implies a highly relativistic outflow of matter and 
electro-magnetic fields lasting on the order of seconds where both external 
and internal shocks are expected (e.g. Rees \& \Mesz, 1992, 1994, Narayan, 
\etal, 1992, \Pacz \& Xu, 1994),  providing a scenario in good qualitative 
agreement with major observational requirements (e.g. \Mesz, \etal, 1994,
\Mesz \& Rees, 1994). Under the cosmological assumption (which we henceforth 
assume) likely energy sources may be, e.g. the coalescence of a compact binary
(Narayan, \etal, 1992) or a failed supernova-like collapse (Woosley, 1993).
While the blast wave propagating into the external medium is highly
relativistic, the reverse shock propagating back into the ejecta is likely
to be only moderately relativistic. In addition, the irregular nature of 
the primary energy release results in the formation of a complex internal 
structure of faster and slower portions of the flow leading to internal
shocks having moderate Lorentz factors ($\sim 1$) in the comoving frame of 
the wind. Hydrodynamic (Waxman \& Piran, 1994) or MHD (Thompson, 1994)
turbulence may be expected in such scenarios and could play a role in the
flow dynamics. In AGN jets also one expects internal shocks, as well as 
termination shocks, and interestingly, the gamma-ray spectrum of blazars
is qualitatively similar to that of GRB. Such systems of shocks and turbulent 
regions provide an environment similar to those thought to lead
to efficient particle acceleration (e.g. Blandford \& Eichler, 1987, 
Jones \& Ellison, 1991, for a review).

In the case of GRB, which we take as a generic example in this paper, 
the shocks are expected to energize an interaction region of spatial 
scale $\Delta \sim c t_{var}\Gamma$ in the wind comoving system, where in the
case of internal shocks $t_{var}$ is the time scale of energy release 
fluctuations (or in the case of the reverse external shock it is the 
light crossing time over the energy deposition region) and $\Gamma$ is the 
mean bulk Lorentz factor of the flow. A multiple-shock structure is likely 
to arise as a result of the energy release fluctutations or the reflection and 
intersection of shocks crossing finite shells or an inhomogeneous outflow. 
Strong, smooth relativistic MHD fluctuations produced by the irregular flow 
motions are expected to arise and interact with particles accelerated in
such shocks.  While details are uncertain, it is possible to consider some
fairly general features of such flow collision regions (FCR), encompassing 
an ensemble of internal shocks and developed turbulent motions, of maximum 
comoving lenght scale $\Delta$ and occuring at a lab frame distance $r_d \sim 
c t_{var} \Gamma^2$ from the center of the event.
Large-amplitude variations of the bulk velocity and the magnetic field can be 
expected on comoving length scales $\l = \alpha \Delta$, with $\alpha < 1$ 
inversely proportional to the number of shocks. For the multiple secondary 
shock generation expected from flow collisions we estimate $\alpha 
\leq 10^{-1}$.    
    
In addition to such larger scale hydrodynamic substructure, we suggest that 
instabilities and nonlinear effects provide some sort of cascading process 
leading also to a wide spectrum of MHD fluctuations on smaller scales. Note 
that small scale MHD motions may have nonrelativistic bulk velocities (unlike
the large scale motions containing most of the energy). Such violent systems 
are favorable sites for nonthermal particle acceleration. The nearest 
nonrelativistic analogue are the corotation interaction regions of the solar 
wind, which have been known for a long time to be an efficient source of low 
energy cosmic rays (e.g. McDonald et al. 1974). Similar systems were suggested 
by Rees (1987) as sites for energetic particle acceleration in AGN jets.    

In the next section we consider the process of nonthermal electron and
positron spectrum formation in FCRs. While the specific examples and numerical 
values refer to cosmological gamma ray bursts, similar considerations apply 
to galactic halo burst models, and may also be of relevance for other
astrophysical (e.g. AGN or galactic jet) sources involving relativistic 
flows and shocks. 
             
\section{{Nonthermal lepton spectra}}

We consider here a nonequelibrium processes of transformation of the 
power in the baryon bulk motions of the relativistic wind to nonthermal 
leptons, and the temporal evolution of the lepton spectra. 
Turbulent plasma motions and shocks have been considered as generic sources
of nonthermal particles since Fermi's pionering work on statistical 
acceleration (e.g. Blandford \& Eichler, 1987). There are three important 
time scales in our problem. The first one is of the order of the cyclotron 
gyration time of relativistic e$^{\pm}$, which is characteristic of the fast 
preacceleration process taking place at shock fronts. The spectra of leptons 
accelerated by single shocks may be steeper than what is needed to explain 
directly gamma-ray burst photon spectra. Thus we consider the individual
shocks as injection agents providing superthermal particles that can be 
further subject to diffusive acceleration through scattering on resonant 
fluctuations and large scale MHD plasma motions. The latter, as shown below, 
can produce very hard lepton spectra on longer, subhydrodynamical and 
hydrodynamical timescales.

\subsection{Cyclotron time scale}

Superthermal particles can be naturally extracted from the thermal pool by 
collisionless shocks, and this process operates on cyclotron timescales. The 
microscopic physics of particle acceleration in relativistic shocks is very 
complicated, and several fundamental aspects remain unclear, although some 
important results highlighting the distinctive features of such shocks have 
been obtained, e.g. Hoshino \etal (1992). These authors find that for 
transverse relativistic shocks pair acceleration to nonthermal energies occurs 
if the upstream flow contains ions carrying most of the energy flux, and they 
speculate that this may extend also to electron-proton plasmas. (Such 
conditions are typical of GRB, where a proton component is expected and pairs 
are present but do not dominate the energy density). Hoshino, \etal (1992)
obtain a downstream nonthermal pair distribution $N \propto \gamma^{-2}$, 
where $N$ is the number of particles within the range $d\gamma$, and their
results indicate that a fraction $0.1-0.2$ of the upstream baryon 
flow energy goes into magnetosonic waves which accelerate the nonthermal pairs.
Microscopic simulations of relativistic quasiparallel shock are not yet 
available, but for nonrelativistic shocks in proton-electron plasmas an 
injection fraction $\zeta \sim 10^{-3}$ is typical (e.g. Giacalone \etal, 
1992). 

\subsection{Subhydrodynamical time scale}

A common attribute of any Fermi-type acceleration process is the 
isotropization of the fast lepton distribution due to the scattering by 
magnetic field fluctuations. From momentum conservation one infers that to 
provide efficient particle scattering the magnetic field fluctuation must 
have an energy density comparable to that of the fast particles in the FCR 
rest frame. Resonant scattering of superthermal leptons will also be 
accompanied by stochastic acceleration with a typical time scale
$\tau^{st}_a \propto  (c/u_{ph})^2 \times (\lambda/c)$, where $u_{ph}$ is 
the phase velocity of the waves resonating with the scattered particle. 
For the mean free path $\lambda(\gamma)$ of a relativistic charged lepton 
within such a system of superthermal particles, the fluctuation spectrum can 
be approximated as being continuous on scales much larger than the electron 
gyroradius, due to strong dissipation. For a broad range of magnetic field 
fluctuations with spectral energy density $W(k)\propto k^{-\mu}$, one expects 
$\lambda \propto \l \times (r_g /\l)^{(2-\mu)}$ from standard quasilinear 
theory (e.g. Blandford \& Eichler, 1987), where $r_g = 1.6\times 10^3~ 
\gamma~ B^{-1}$ cm is electron gyroradius. The use of quasilinear theory 
for the effects of fluctuations on scales $\leq \l$ is justified because of 
the relatively small amplitudes of the resonant fluctuations (see also 
Hoshino \etal, 1992) in the sub-hydrodynamic MHD regime. Thus here we 
calculate the temporal evolution of the $e^{\pm}$ spectrum using a standard 
Fokker-Planck treatment.

This process leads to a nonthermal electron (or $e^\pm$) spectrum of the 
form $N(\gamma) \propto \zeta n \gamma^{1 - \mu}$, which develops on a
sub-hydrodynamic time scale of the order of a few $\tau_a^{st}$, which
for $u_{ph}\sim c$ is
\begin{equation}
\tau^{st}_a \sim (r_g/l)^{2-\mu} \times l/c ~.
\end{equation}
Typically $(r_g/l)^{2-\mu} \ll 1$ for the energies of interest (except for 
the case $\mu$ = 2). From energy conseravtion, this energy spectrum extends 
up to 
\begin{equation}
\gamma_{\ast} \sim 
\left[\gamma_i~m_p/m_e~\epsilon~ \zeta^{-1}\right]^{1/(3 - \mu)}~,
\end{equation}
where $\gamma_i$ is the initial proton (lepton) Lorentz factor,
$\epsilon < 1$ is the portion of the total upstream power in baryons  
pumped into turbulent fluctuations, $\zeta$ is the lepton injection fraction.
The difference between the steeper $\gamma^{-2}$ spectrum injected near the
shock and the harder $\gamma^{1 - \mu}$ spectrum produced by the MHD 
fluctuations arises because of the different spatial extent of the 
acceleration regions. In the former the acceleration and escape time scales 
are comparable, $t_a\sim \kappa/u^2$, where $u$ is flow velocity and $\kappa
\sim v\lambda$ is diffusion coefficient with $v\sim c$ the particle velocity, 
while $t_{esc}\sim \delta^2/\kappa$ where $\delta\sim (v/u)\lambda \sim 
\kappa/u$ is the width of particle acceleration region near the shock, so 
$t_a^{-1}~t_{esc} \sim u^2 \delta^2/\kappa^2 \sim 1$. For scattering by MHD 
fluctuations between the shocks, however, the acceleration time is much shorter 
than the escape time: $t_a \sim (r_g/\l)^{2-\mu}(\l/c) \sim \lambda/c$, while 
$t_{esc}\sim\Delta^2/\kappa\sim (\Delta/\lambda)^2 (\lambda/c)\gg (\lambda/c)$.

Since typical MHD turbulent spectra have indices $\mu$ with $1.5\leq\mu\leq 2$, 
the resulting particle spectra are $\propto \gamma^{-1}$ or flatter. We 
conclude that a substantial portion of the power in turbulence and some 
portion $\sim \epsilon$ of the upstream baryon power in the flow is 
transferred to the charged leptons at energies near $\gamma_{\ast}$. 
An estimate of $\epsilon$ and $\zeta$ would depend on uncertain details 
such as the ratio of the matter to antimatter content in the flow, wave
modes supported by the plamsa, etc. In the absence of other information, we 
will adopt here $\zeta\sim 10^{-3}$ and $\epsilon \sim 0.1$, compatible with 
the numerical results of Giacalone, \etal (1992) and Hoshino, \etal (1992).
Taking as an example $\mu=1.5$, an estimate of the break energy for the above 
typical parameters of the model gives $\gamma_{\ast} \sim 3\times 10^3$.

Synchrotron losses of relativistic pairs have a time scale $\tau_{syn} \approx 
5\times 10^8 B^{-2}\gamma^{-1}$ s, if B is measured in G. Thus for the
formation of a hard branch of the pair spectrum at least up to the Lorentz 
factor $\gamma_\ast$ due to resonant acceleration the scale $\l$ must satisfy 
the condition $\l \leq 10^{(3\mu + 12)/(\mu -1)}~ B^{\mu/(1-\mu)}~
\gamma^{(\mu-3)/(\mu-1)}$ cm. This imples, for $B = 10^4$ G and $\gamma_{\ast} 
\sim 1-2\times 10^3$ the scale $\l \leq 10^7$ cm if turbulent fluctuations 
have index $\mu$ =2, and $\l \leq 10^{12}$ cm if $\mu$ = 1.5. 

On the subhydrodynamical time scale the distribution of nonthermal pairs for  
$\gamma \gg \gamma_{\ast}$ will be highly intermittent, with nonthermal pairs 
at these energes concentrated in the  vicinity of the shocks, since they 
lose their energy before being mixed within the FCR. The synchrotron photon 
spectrum of the system beyond the break might be dominated by the brightest 
spots from some particular shock or the superposition of contributions from 
a few shocks. The resulting spectral shape just beyond the break may be 
rather complicated, reflecting with some modifications the injection spectrum.

\subsection{Hydrodynamical time scales}

In addition to the above effects, one can also expect acceleration from 
processes occuring on the longer hydrodynamical comoving time scales of the 
order of $\l/c$. The electric fields induced by turbulent motions of plasmas 
carrying magnetic fields on different scales lead to statistical energy gains 
of the superthermal charged particles. For nonrelativistic MHD turbulence the 
particle energy change over a turbulent correlation length (or correlation 
time) is small, because the induced electric field is smaller then the 
entrained magnetic field. However, the distinctive feature of statistical 
acceleration in the {\it relativistic} MHD turbulence and shocks on larger 
scales expected in the FCR, is the possibility of a substantial particle energy 
change over one correlation scale, because the induced electric fields are 
no longer small. In this case a Fokker-Planck approach cannot be used.
Instead, we argue here that it is possible to calculate the energy spectra 
of nonthermal particles within FCRs  using  an integro-differential equation 
which is a generalization of the Fokker-Planck approach (for details see the 
review by Bykov \& Toptygin, 1993, hereafter BT). 

Consider charged test particles interacting with a wide spectrum of MHD 
fields and an internal shock ensemble produced by the colliding flows within 
generalized FCRs. In the wind comoving frame, we can assume the fluctuations 
on all scales up to $\sim \Delta$ (including the internal shock ensemble) to
be nearly isotropic (for the latter, it is enough if they are forward-backward 
symmetric). The small mean free path $\lambda$ of the superthermal particles 
leads to their isotropy in the frame of the local bulk velocity fluctuations. 
The assumed statistical isotropy of the bulk velocity fluctuations in the 
comoving frame of the wind results then in a nearly isotropical particle 
distribution, after averaging over the ensemble of internal shocks and 
accompanying motions on scales $\sim \l$.

To calculate the spectrum of nonthermal leptons accelerated by the ensemble of 
internal shocks and large-scale plasma motions in the FCR (averaged over the 
statistical ensemble of large scale motions) we use a kinetic equation for 
the nearly-isotropic distribution function $N=\gamma^2 F$, which takes into 
account the non-Fokker-Planck behavior of the system (see Bykov 1991 and BT),
\begin{eqnarray}
\frac {\partial F({\rm r},\xi,t)}{\partial t} = &  Q_i(\xi) +   
  \int_{-\infty}^{\infty} 
{\rm d} \xi_1 \; D_1 (\xi ~-~\xi_1 )\;\Delta F({\rm r},\xi_1,t) \nonumber\\
 & + \left(\frac {\partial^2 }{\partial \xi^2} + 
    3\frac {\partial }{\partial \xi} \right)\int_{-\infty}^{\infty}
   {\rm d}\xi_1\;D_2(\xi~-~\xi_1)\:F({\rm r},\xi_1,t) 
\end{eqnarray}
Here $\xi =\ln(\gamma/\gamma_i)$, $\gamma_i$ is the Lorentz factor of the 
injected particles, $Q_i(\xi) \propto \zeta c n \l^2$ is the rate of nonthermal 
particle injection, $n$ is the lepton number density in the FCR comoving frame. 
The kernels of the integral equation Eq.(3)
determining the spatial and momentum diffusion are expressed through 
correlation functions describing the statistical properties of the large scale 
MHD turbulence and shock ensemble. Following the renormalization method, the 
Fourier transforms of the kernels $D_1^F(s)$ and $D_2^F(s)$ are solutions of 
a transcendental algebraic system of equations of the form  $D_{1,2}^F = 
\Phi_{1,2} (D_1^F, D_2^F,s)$. Here $s$ is a variable which is Fourier 
conjugate of $\xi$. Equation (3) and the renormalization equations are valid
only for particles with sufficiently small mean free paths $\lambda(\gamma) 
\ll \Delta$. 

The crucial point is that the solution of equation (3) has a universal
behavior, only weakly dependent on the complicated details of the turbulent 
system. The stationary solution to Eq.(3) with a monoenergetic injection rate 
$Q_i$ has an asymptotical behavior of a power-law form, $N \propto Q_i 
\gamma^{-\sigma}$ (Bykov, 1991), where $\sigma = - 0.5 + [2.25 + \theta 
D_1(0)D_2^{-1}(0)]^{0.5}$, and we took $\theta \sim (\l/\Delta)^2$. For 
conditions typical of developed turbulence, the ratio of the rate of the 
scattering to the acceleration rate is $D_1(0)D_2^{-1}(0) < 1$ (see BT), and
for $\theta <1 $ we obtain $\sigma \sim 1$. This hard $\gamma^{-1}$ 
spectral behavior arises because the acceleration time $\tau_a\sim \l/c \sim
\alpha\Delta/c$ is much shorter than the escape time at the relevant energies, 
$\tau_{esc}\sim\Delta^2/\kappa\sim\Delta^2/(\l c) \sim \Delta/(\alpha c)$. 
The power needed to produce such a spectrum of nonthermal particles increases 
$\propto \gamma_{max}$, so it is important to understand its temporal 
evolution. 

In the test particle limit, where the backreaction of the accelerated leptons
on the energy-containing bulk motions  is negligible, we have $N(\gamma,t) 
\propto \zeta n \gamma^{-1}$ for $\gamma \leq \gamma_{\star}(t)$, where 
$ \gamma_{\star}(t) =  \gamma_{i} \exp(t/\tau^h_a)$ and 
\begin{equation}
\tau^h_a \propto \l/c \sim \alpha (\Delta /c)~,
\end{equation}
is the typical hydrodynamical acceleration timescale (see e.g. BT), with 
$\gamma_{i}\sim$ few, $\alpha < 1$ and $\Delta$ the comoving width of the 
region energized by shocks. From the energy balance equation, when the value
$\gamma_{\star}(t) \sim \gamma_i~ m_p/m_e~\epsilon~\zeta^{-1}$ is reached 
the growth must saturate, and the resulting spectrum consists of two branches.
One is the hard spectrum $N(\gamma)  \sim \zeta n \gamma^{-1}$, for 
$\gamma \leq \gamma_{\star}$, where
\begin{equation}
\gamma_{\star} \sim \gamma_i~m_p/m_e~ \epsilon~\zeta^{-1}~. 
\end{equation}
For the typical values of our problem $\gamma_i \sim 1$ and $\zeta\sim 10^{-3}$ 
so $\gamma_\star\sim 10^{5}$ (but it could be even larger since $\epsilon 
\sim 1$ for large scale plasma motions).

\section{ Electron Energization Efficiency}

We outline here the application of the above acceleration scenario to the 
dissipative fireball model of gamma-ray bursts. It has been argued in the 
introduction that one can expect violent flow collision regions (FCRs) to form 
in the dissipative portion of the fireball evolution. We do not go here into a 
detailed discussion of the radiation physics nor do we attempt to model GRB in
detail, concentrating rather on broadly generic examples using typical values 
of the relevant physical quantities.

In the fireball wind models, FCRs might occur around radii $r\sim c t_{var}
\Gamma^2 \sim 10^{12}- 10^{13}$ cm with bulk Lorentz factors $\Gamma\sim 10^2$ 
and mean comoving field strengths $B \sim 10^4 B_4$ G (e.g. Rees \& \Mesz, 
1994). We assume the acceleration to occur beyond the region where significant 
pair formation is expected, i.e.  outside the photosphere, and take as 
numerical examples a lepton injection fraction $\zeta\sim 10^{-3}$, turbulence 
energy fraction $\epsilon \sim 10^{-1}$, and initial injection Lorentz factor 
$\gamma_i \sim 1$. We assume also a broad spectrum of MHD or whistler type 
fluctuations with index $\mu$ = 1.5. The characteristic time scale of the fast 
injection process of leptons is a few microseconds in the FCR comoving frame. 
Then, on a subhydrodynamical time scale $\tau_a^{st}\sim 10 c^{-1}(l r_g)^{1/2} 
\sim  10 c^{-1}({\alpha c t_{var} \Gamma r_g})^{1/2}$ a hard lepton spectrum 
is established, with $N(\gamma) \propto \zeta n \gamma^{-0.5}$ for $\gamma 
\leq\gamma_{\ast}\sim\left[\gamma_i~m_p/m_e~\epsilon~ \zeta^{-1}\right]^{2/3}$. 
The comoving frame time $\tau_a^{st} \siml$ few ms is enough to transfer a
fraction $\sim \epsilon$ of the baryonic power to the accelerated leptons, with 
Lorentz factors $\gamma\sim \gamma_\ast \sim 2\times 10^3$. For steeper 
turbulent fluctuation spectra ($\mu \simg 1.5$) the timescale is somewhat 
longer but still compatible with observational requirements. As they are 
accelerated, the leptons radiate a synchrotron spectrum of the form $\nu F_\nu 
\sim \nu^{1.25}$ up to a break near 0.1 MeV for the parameters used, and 
$\propto \nu^{k}$ above that with $k$ between 0 and -1 if rather weak 
shocks dominate in the FCR.  

Nonthermal lepton acceleration on the longer, hydrodynamical timescales
might be important in GRB external shocks around radii $r\sim 10^{16}$ cm, 
e.g. \Mesz, Rees \& Papathanassiou, 1994. Typical magnetic fields in the 
reverse shock could be $\sim 10$ G, while the hydrodynamic time scale is
$\tau_a^h \sim \alpha \Delta/c \sim$  tens of seconds in the FCR comoving 
frame (and $\Gamma^{-1}$ times shorter in the lab frame). This is enough to 
form a pair spectrum $N(\gamma) \propto \zeta n \gamma^{-1}$ up to Lorentz 
factors $\gamma_{\star} \sim 2\times 10^5$, for these parameters, and a 
synchrotron spectrum with $\nu F_\nu \sim \nu$ and a peak energy near the 
0.5 MeV range.

Recent GRO observations of blazars (McNaron-Brown \etal, 1995) show clear
evidence for broken power-law spectra peaked in the MeV range with a shape
similar to gamma-ray burst spectra (e.g. Greiner, \etal, 1995). From the
similarity of the acceleration scenarios expected in both, involving 
relativistic shocks and turbulence, one may speculate on the possible 
applicability of the above processes to explain blazar spectra.

In principle, an injected fraction $\zeta <1$ of protons may also be 
accelerated by the same mechanisms.  The maximum proton Lorentz factors would 
in this case be $\gamma_{p,\ast} \sim (\epsilon \gamma_i/\zeta)^{1/(3-\mu)}$ 
or $\gamma_{p,\star}\sim (\epsilon \gamma_i/\zeta)$, which for $\epsilon\sim 
10^{-1},~\zeta\sim 10^{-3}$ could be as high as $10^4$.
However, the fraction of postshock proton energy going into such a
flat, nonthermal relativistic proton component is at most $\epsilon$, 
comparable with the fraction of energy going into the nonthermal flat 
lepton spectrum.

The efficiency of the transfer of energy from the proton to the lepton
component in these models is high, typically of order $e_{pe}\sim
(\zeta/\gamma_i)(m_e /m_p ) \gamma_e^{3-\mu} \siml \epsilon$ for 
subhydrodynamic, and $e_{pe} \sim (\zeta /\gamma_i)(m_e/m_p) \gamma_e \siml 
\epsilon$ for hydrodynamic acceleration, where $\gamma_e$ can go up to 
$\gamma_\ast$ or $\gamma_\star$, and $\epsilon\siml 1$ is the fraction of 
upstream proton energy converted into turbulence in the semirelativistic wind 
and reverse blast wave shocks, $\gamma_i$ is initial Lorentz factor and 
$\zeta$ is lepton injection fraction. This high efficiency is due to the very 
hard lepton spectra achieved with $t_{acc}\ll t_{esc}$, $N(\gamma) \propto 
\gamma^{-1}$ or flatter, which puts most of their energy near the upper break 
value $\gamma_\ast$ or $\gamma_\star$.This is a significant fraction of the 
equipartition value between the accelerated leptons and the bulk of the 
shocked protons. Similar shocks and turbulent regions are likely to be present 
in AGN or galactic jets. If these are electron-proton jets, as opposed to 
electron-positron jets, most of the energy is in the protons (as for the 
GRB case), and a significant fraction of it should be channeled into the 
electrons. Since the leptons, due to their smaller mass, are responsible 
for most of the radiation, this mechanism of proton-electron energy sharing 
fulfills a major prerequisite for a high radiative efficiency in GRB and 
other nonthermal gamma-ray sources.

\bigskip

{\it Acknowledgements}: We are grateful to the Institute for Theoretical 
Physics, University of California, for its hospitality, and to participants 
in the ITP Workshop on Nonthermal Gamma-Ray Sources for discussions.
This research is supported through NSF PHY94-07194, NASA NAG5-2857 and the
International Science Foundation (grants NU 3000, 3300) and Russian BRF
(grant 95-02-04143a).

\end{document}